\documentclass[prl, amsfonts, twocolumn, nofootinbib, showpacs, letter]{revtex4}
\usepackage{graphicx, epsfig}

\begin{document}
\title{Screening and Antiscreening of the MOND field in Perturbed Spherical Systems}
\author{Reijiro Matsuo$^{1}$ and Glenn Starkman$^{1}$}
\affiliation{$^{1}$CERCA, Department of Physics, Case Western
Reserve University, Cleveland, OH~~44106-7079}


\begin{abstract}

 \widetext
In the context of Modified Newtonian Dynamics (MOND), we study how perturbation of a spherically symmetric system would affect the gravitational field. In particular, we study systems of perturbed and unperturbed spherical shells.  For a single perturbed shell, we show that the field inside the shell is much smaller than what would be expected from a naive scaling formula. The strength of the perturbation field within the shell is screened by the spherically symmetric component of the mass, and is reduced as the spherically symmetric component is increased.  For a two-shell system, we again show that the perturbed field is screened by the shells, no matter which shell's mass distribution is perturbed. The field within the inner shell is most suppressed when the inner  and outer shells coincide.  However, for a very light inner shell, the perturbation to the field can be enhanced.  The enhancement is typically larger for smaller inner shells, and the perturbed field can be amplified  by almost a factor of 2.  The relevance to the  effect of external fields on  galaxy dynamics is discussed.
\end{abstract}

\maketitle

\section{\bf Introduction}
\indent
Most of the astrophysical computation concerns a local system embedded in much larger background system.  For example, Solar system is embedded in Milky Way, a galaxy in a cluster, a cluster to the supercluster etc.  In GR or Newtonian gravity, the external field effect on the local system due to much larger background mass density are often neglected based on two simplifying assumptions: that the background mass distribution is spherically symmetric and/or the background distribution is located very far from the local system.  Under the first assumption, one would invoke Birkhoff's theorem that the field inside a spherical cavity should vanish.  (Inclusion of the local mass distribution may spoil the spherical symmetry of the whole system, and in GR, the background mass distribution may exert force on the local system.  However the external field induced in this way is negligibly small.  See \cite{Dai}).  Under the second assumption, the inverse-square law will suppress the contribution of the portion of the background mass distribution that is  far from the local system.  Therefore, even when the background mass distribution is only approximately spherically symmetric, the external field effect in GR or Newtonian gravity would not significantly alter the local dynamics.

It might be naively expected that the external field effect is more significant in MOND theory.   A small perturbation of the background mass distribution from spherical symmetry might be expected to alter the internal field configuration significantly.  This is because the field of a bounded mass distribution scales as $\sim \frac{1}{r}$ in MOND, and the field due to the aspherical part may therefore persist to greater distances.  Of course, the modified Possion equation in MOND is non-linear, and it does not embody the superposition principle.  One cannot treat the contribution from the spherically symmetric part of the mass distribution and the aspherical part separately.  
The total mass distribution has to be considered in order to infer the size of any external field effect. 
The external field effect in MOND has been receiving growing interest \cite{Bruneton,WW}.  It is argued that MOND theory is consistent with the observed escape velocity in the solar neighborhood if the Milky Way is embedded in a constant external field of $\sim 0.01a_{0}$.  (Here $a_0$ is
the universal critical acceleration that characterizes MOND, $a_{0}\simeq 1.2\times10^{-8}{\mathrm cm}/{\mathrm s}^2$.
 On the other hand, Wu et al \cite{Wu2} has showed that an external field of approximately $0.03a_{0}$ is needed to bind the Large Magellanic Cloud  to the Milky Way.  Since the external field effect on the local escape speed presents a strict test of MOND theory, it is of interest to understand when, if at all, the aspherical perturbation of very distant backgrounds could be neglected.  
 
 In an earlier work \cite{Dai}, we studied  the fields inside spherical shells as probed by non-negligible test masses.
 We showed that when those test masses are placed off-center, thereby breaking the spherical symmetry,
 the acceleration of the test mass can become comparable to the (external) surface gravity of the shell.
 In what follows, we consider  simplified systems of spherical shells with perturbations in order to better 
 understand how the perturbations of spherically symmetric  mass distributions translate into perturbed
 gravitational fields.

\section{MOND}
MOND \cite{MOND1,MOND2,Milgrom1} is an alternative to the canonical dark-matter scenario in which the gravitational force law 
(or Newton's second law) is modified to account for the missing gravitational field. In 1984, Bekenstein and Milgrom introduced the Lagrangian formulation of Modified Newtonian dynamics (MOND) \cite{Bekenstein}. 
The field equation of MOND is derived from the Lagrangian
\begin{equation}
L=-\int d^{3}r\left\{\rho\psi+(8\pi G)^{-1}a_{0}^{2}\mathcal{F}\left[\frac{(\nabla\psi)^{2}}{a_{0}^{2}}\right]\right\} \quad , \label{Lagrangian}
\end{equation}
where $\psi$ is the gravitational potential.
$\mathcal{F}(y^{2})$, with $y\equiv|\nabla\psi|/a_{0}$, is an arbitrary universal function,
that together with $a_{0}$, the characteristic scale of MOND,  specifies the theory.
Varing L with respect to $\psi$ yields a modified Possion equation:
\begin{equation}
\nabla\cdot[\mu(|\nabla\psi|/a_{0})\nabla\psi]=4\pi G\rho(\mathit{|\mathbf{r}|}) \quad , \label{ModPossion}
\end{equation}
where $\mu(y)\equiv\mathcal{F}'(y^{2})$. 
$\mu(y)$ must approach 1 as $|y|\gg 1$ and $|y|$ as $|y|\ll 1$,
in order that the field scale as $\frac{1}{r^{2}}$ near a spherical  mass distribution (the usual Newtonian result)
and as $\frac{1}{r}$ far from the mass distribution to explain flat galaxy rotation curves. 

One commonly used form of $\mu$ is
\begin{equation}
\mu(y)=\frac{|y|}{\sqrt{1+|y|^{2}}} \label{mufunction}\quad .
\end{equation}
(However see \cite{Famaey} for different form of $\mu$ function.) 
The value of $a_{0}$ is then given by phenomenological fit. 
We will adapt the value derived by Begeman et al \cite{Begeman}
in the study of external galaxies with high quality rotation curves.
\begin{equation}
a_{0}=1.2\times10^{-10}m/s^{2}\quad . \nonumber
\end{equation}

For a bounded mass distribution of total mass $M$, we define a transition radius
\begin{equation}
R_{t}=\sqrt{GM/a_{0}} \quad .
\end{equation}
$R_{t}$ indicates a point at which the Newtonian field approximately equals $a_{0}$,
and this is about the point at which the field switches from Newtonian $1/r^2$ to MOND's $1/r$.

We now define three quatntities, MOND acceleration $\mathbf{g}_{M}$, Newtonian acceleration $\mathbf{g}_{N}$ and 'naive' MOND acceleration $\mathbf{g}_{s}$.  MOND acceleration comes from the full solution of the modified Poisson equation \ref{ModPossion}, and is defined to be
\begin{equation}
\mathbf{g}_{M}=-\nabla\psi
\end{equation}
Newtonian potential $\phi$ is the soloution of a ordinary Poisson equation
\begin{equation}
\nabla\cdot\nabla\phi_{N}=4\pi G\rho(\mathit{|\mathbf{r}|}),
\end{equation}
and Newtonian acceleration $\mathbf{g}_{N}$ is defined to be
\begin{equation}
\mathbf{g}_{N}=-\nabla\phi_{N}.
\end{equation}
From the form of the modified Poisson equation, 'naive' MOND acceleration $\mathbf{g}_{s}$ and Newtonian acceleration $\mathbf{g}_{N}$ for a given density distribution can be related by the algebraic relation
\begin{equation}
\mu(|\mathbf{g}_{s}|/a_{0})\mathbf{g}_{s}=\mathbf{g}_{N}
\end{equation}
This is the original formation of MOND and sometimes incorporated to calculate the galactic rotation curves \cite{Kent,Milgrom2,Begeman}.
Inverting the relation we will find
\begin{equation}
\mathbf{g}_{s}=\nu(|\mathbf{g}_{N}|/a_{0}) \label{naive}
\end{equation} 
where $\nu(x)=I^{-1}(x)x$ and $I(x)=x\mu(x)$.  Usually $\mathbf{g}_{s}$ is not equal to $\mathbf{g}_{M}$ since $\mathbf{g}_{s}$ in general is not curl-free. ($\mathbf{g}_{M}$ by defnition has to be curl-free since it is gradiant of a potential).  Important exception is if $\mathbf{g}_{N}$ only depends on one parameter, which is the case for spherical, cylindrical or plane symmetry, then $\mathbf{g}_{M}=\mathbf{g}_{s}$ \cite{Brada}.  Then if the perturbation is small, we might expect that MOND acceleration can well be approximated by $\mathbf{g}_{s}$.
\section{Code} 
The numerical code used in this paper is developed by Milgrom \cite{Milgrom}. It is the same code we used in our earlier paper \cite{Dai}, and it is the implementation of Milgrom's code on a spherical lattice.  We set the number of angular glid N and that of radial glid L to be 120 and 400 respectively. $N=120$ tlanslates into the angular resolution of $1.5^{\circ}$.       
\section{Perturbative calculation}
\begin{figure}
\centering{
\includegraphics[width=3.0in, height=3.3in]{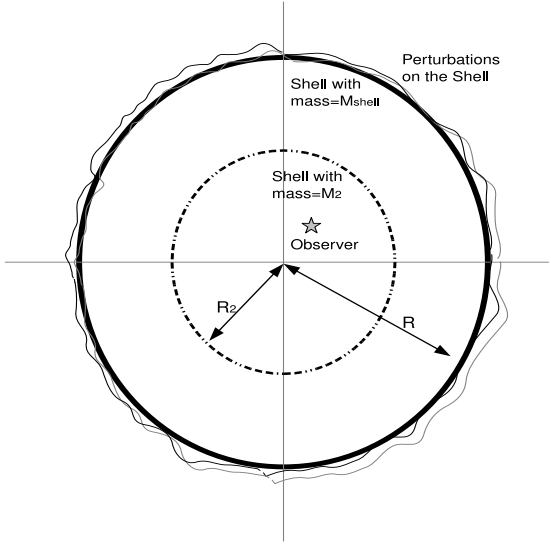}}
\caption{This figure shows the mass distribution of the system.  A shell with $M_{sh}=2.0\times10^{13}M_{\odot}$ and $R=0.64 Mpc$ is placed concentric with the origin of the coordinate .  A massless observer is located at $r=0.05R$ with an angular displacement of $45^{\circ}$ from z-axis.  An azimuthally symmetric perturbation is placed on the shell.  For the multiple shell system, a second shell with $mass=M_{2}$ with $radius=R_{2}$ will be placed concentric with the first shell.}
\label{dist}
\end{figure}
We first consider a small aspherical perturbation on the shell as in FIG. \ref{dist}.  The center of the shell is taken to be the conter of the spherical coordinate for the sake of simplicity. The radius of outer shell and its mass is fixed to $R=0.64 Mpc$ and $M_{sh}=2.0\times10^{13}M_{\odot}$ which would be a typical parameters for a gas-rich cluster \cite{Angus}.  
Let the density distribution of the perturbation and the spherical shell be $\rho_{l}$ and $\rho_{sh}$, respectively, then we will write
\begin{eqnarray}
\rho_{sh}&=&\sigma_{sh}\delta(r-R)  \quad \nonumber \\
\rho_{l}&=&\epsilon\sigma_{sh}Y(\theta)\delta(r-R),
\end{eqnarray}
where $\sigma_{sh}$ is surface mass density for the shell, and $Y(\theta)$ is some angular function. 
The unperturbed field $\psi_{0}$ satisfies the modified Poisson equation:
\begin{equation}
\nabla\cdot[\mu(|\nabla\psi_{0}|/a_{0})\nabla\psi_{0}]=4\pi G\rho_{sh} \quad .\nonumber
\end{equation}
The exact solution for $\nabla\psi_{0}$ can be found by applying Gauss's theorem. 
In terms of the quantity $u=\frac{GM_{sh}}{a_{0}r^{2}}$,
\begin{eqnarray}
\nabla\psi_{0}&=&-a_{0}\sqrt{\frac{1}{2}u^{2}+\sqrt{\frac{1}{4}u^{4}+u^{2}}}\,\hat{r} \\
                                   &\simeq-&a_{0}u^{\frac{1}{2}}(1+\frac{1}{4}u+...)\hat{r} \quad .                   \nonumber
\end{eqnarray}
for exterior of the shell, and
\begin{equation}
\nabla\psi_{0}=0 \nonumber
\end{equation}
for the interior of the shell.
The expansion in small $u$ is valid for $r\gg\sqrt{\frac{GM_{k}}{a_{0}}}$. 
The solution $\psi$ will satisfies the following jump conditions at r=R:
\begin{eqnarray}
\mu\left(\frac{|\nabla\psi_{out}|}{a_{0}}\right)\nabla\psi_{out}\cdot\hat{r}|_{r=R}&-&\mu\left(\frac{|\nabla\psi_{in}|}{a_{0}}\right)\nabla\psi_{in}\cdot\hat{r}|_{r=R} \quad \nonumber \\
                                                                                   &=&4\pi G[\sigma_{sh}+\epsilon\sigma_{sh}Y(\theta)]  \nonumber \\
                                             \nabla\psi_{out \|}-\nabla\psi_{in \|}&=&0 
\end{eqnarray}
Exterior  to the shell, the spherically symmetric solution $\nabla\psi_{0}$ will dominate, and one can expand the modified Poisson equation around
$\psi_{0}$. If the small correction to the potential is denoted by $\psi_{1}$, one can write the expansion as 
\begin{eqnarray}
\nabla\cdot\left[\mu\left(\frac{|\nabla\psi|}{a_{0}}\right)\nabla\psi\right]=\nabla\cdot[\mu(x_{0})\nabla\psi_{0}] \quad\quad\quad\quad\nonumber \\
                                              +\nabla\cdot\left[[[\hat{r}\cdot\nabla\psi_{1}]L(x_{0})\hat{r}+\nabla\psi_{1}]\mu(x_{0})\right] 
                                              + h.o.
\end{eqnarray}
Here $L$ is a function given by $\frac{x\mu^{'}(x)}{\mu(x)}$, and $x_{0}$ is given by $x_{0}=\frac{|\nabla\psi_{0}|}{a_{0}}$.
In the interior of the shell, the field is caused by aspherical perturbation, and one may assume that the 'simple' form of $\mu$ function is applicable. (i.e. $\mu(x)\sim x$ as is the case for the system in the deep MOND regime).  
Then $\psi_{1}$ has to satisfy the following equations  interior and exterior to the shell. 
\begin{eqnarray}
\frac{1}{r^{2}}\frac{\partial}{\partial r}\left[r^{2}[1+L(x_{0})]\mu(x_{0})\frac{\partial\psi_{1}}{\partial r}\right]-\frac{\mu_{0}}{r^{2}}\mathcal{L}^{2}\psi_{1}&=&0 \,\, \mbox{for $r>R$}  \nonumber \\
\nabla\cdot\left[\frac{|\nabla\psi_{1}|}{a_{0}}\nabla\psi_{1}\right]&=&0 \,\,\mbox{for $r<R$} \nonumber \\
 \label{simplemu} 
\end{eqnarray}
where $\mathcal{L}^{2}$ is the angular-momentum-squared operator.
At $r=R$, the jump conditions for $\rho_{l}$ should be satisfied:
\begin{eqnarray}
[1+L(x_{0})]\mu(x_{0})\frac{\partial\psi_{1 out}}{\partial r}|_{r=R} \hspace{1.5in} \\
\hspace{1.0in}-\frac{|\nabla\psi_{1 in}|}{a_{0}}\frac{\partial\psi_{1 in}}{\partial r}|_{r=R}=4\pi G\sigma_{sh}Y(\theta)\nonumber \\
\frac{\partial\psi_{1 out}}{\partial\theta}|_{r=R}-\frac{\partial\psi_{1 in}}{\partial\theta}|_{r=R}=0\hspace{0.6in}\nonumber
\label{jump}
\end{eqnarray}
Let us now consider a particular form of the perturbation
\begin{equation}
\rho_{l}=-\epsilon\sigma_{sh}P_{l}[cos(\theta)]\delta(r-R) \label{perturbation}
\end{equation}
where $P_{l}$ is the Legendre Polynomial of lth order.
The second of equation (\ref{simplemu}) belongs to a class of equations known as p-Laplacian for $p=3$ and $N=3$, where N is the number of Euclidian dimensions.  (For p-Laplace equation, see for example \cite{Lindeqvist}.  For the special case of p=N, see \cite{Rouba}).  Little is known about its solution.  However, if the angular dependence of $\psi_{1}$ is proportional to $cos(\theta)$, by observation one can find that $rcos(\theta)$ is a solution of equation (\ref{simplemu}) inside the shell.  Then for the particular form of the perturbation with $l=1$, the matching of the solution can be performed for inside and outside of the shell.
\begin{equation}
\psi_{1}=\left\{ \begin{array}{c}\Psi_{sh}\epsilon cos(\theta)\left[\frac{R}{2r_{t}}-\frac{5}{56}\left(\frac{r_{t}}{R}\right)+...\right]\left[\frac{r_{t}}{r}+\left(\frac{11}{28}\frac{r_{t}}{r}\right)^{3}+... \right] \\
\hspace{2.2in} \mbox{for $r> R$}\\  
\Psi_{sh}\epsilon cos(\theta)\left[\frac{r}{2R}+\frac{6}{56}\left(\frac{r_{t}}{R}\right)^{2}\frac{r}{R}\right] \hspace{1.3in}\\
\hspace{2.2in} \mbox{for $r\leq R$} \end{array} \right. \label{potential}\quad  
\end{equation} 
where $\Psi_{sh}=\sqrt{GM_{sh}a_{0}}$.
\begin{figure*}[t!]
\centering{
(a)\includegraphics[width=3.0in]{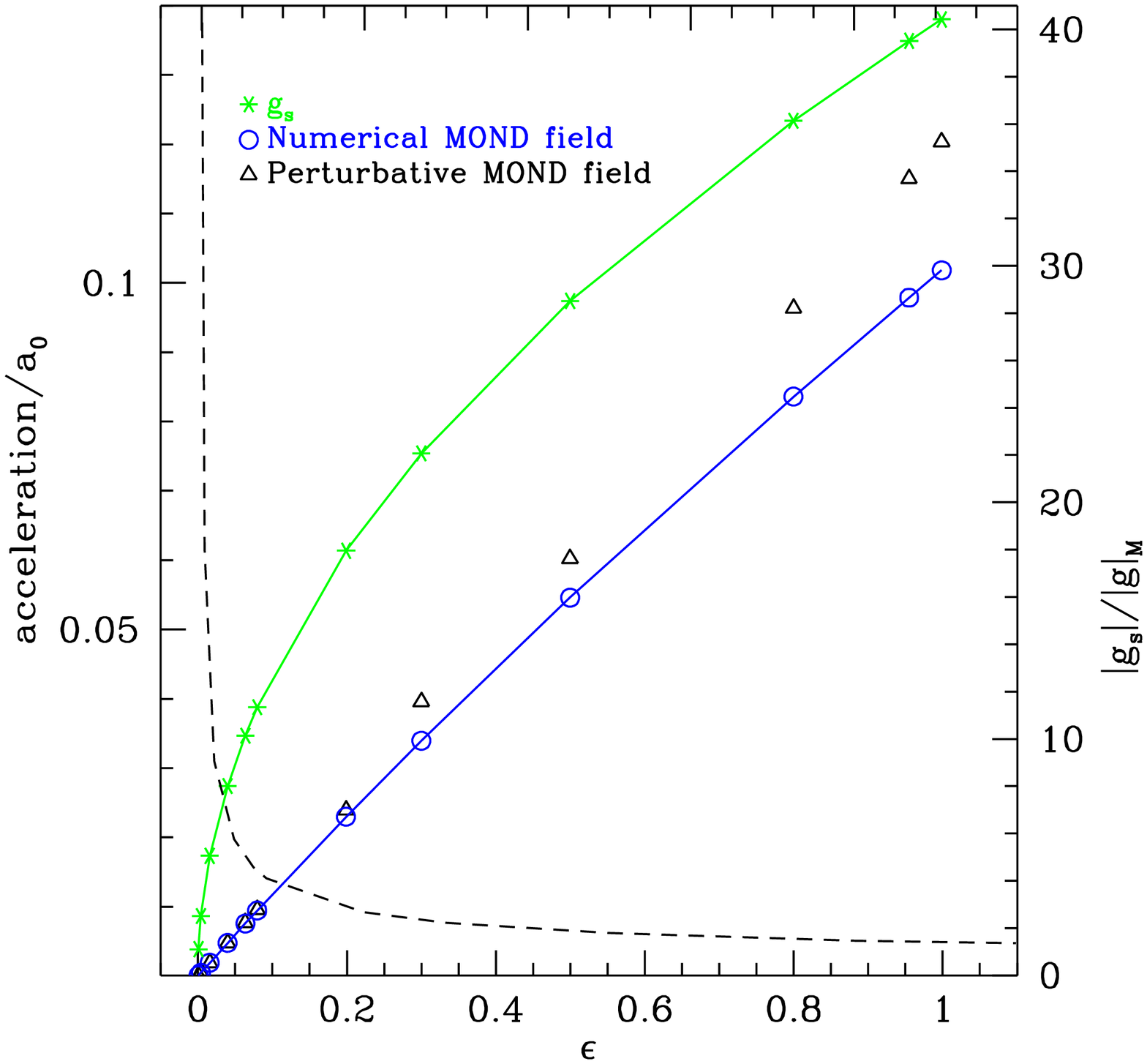}
(b)\includegraphics[width=3.0in]{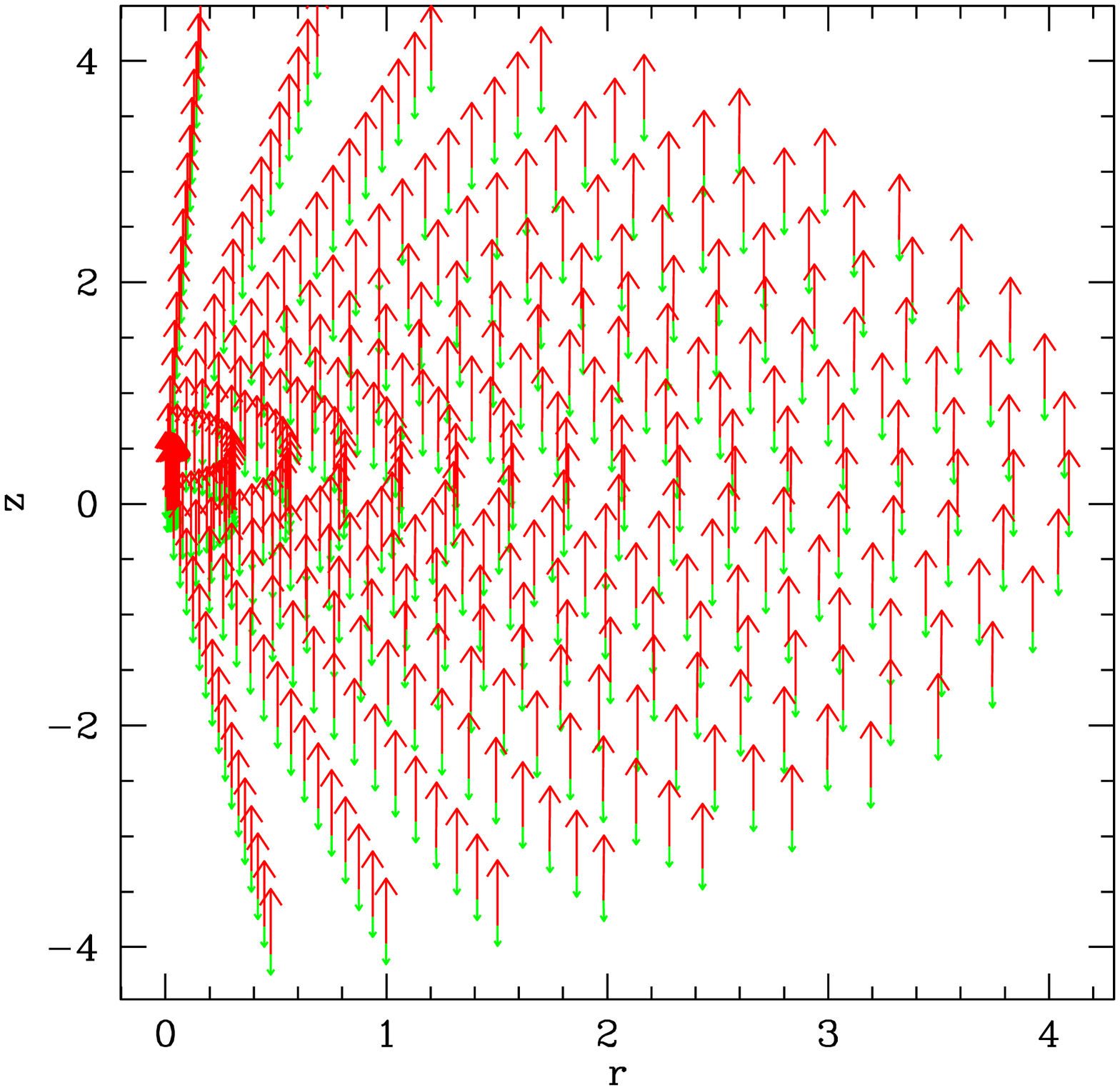}}
\caption{(a) This figure plots the field at a particular point  ($r=0.05R$ and $\theta=45^{\circ}$) inside a shell
with an $l=1$ perturbation  of the form given by equation (\ref{perturbation},
as a function of the strength $\epsilon$ of the perturbation.  The strength of the numerical MOND field $\mathbf{g}_{M}$, 
the perturbative MOND field and the naive MOND field $|\mathbf{g}_{s}|$ are shown. 
The dashed line indicates the ratio of $|\mathbf{g}_{s}|$ to $|\mathbf{g}_{M}|$ with the scale given on the right axis. 
 (b)This figure shows the field configuration inside the shell for a perturbation of order $l=1$ with $\epsilon=0.1$.  
 Green arrows indicate the numerical MOND field, and red arrows indicate the difference between the numerically calculated 
 MOND field and the naive MOND field, $\mathbf{g}_{M}-\mathbf{g}_{s}$.  The sizes of the arrows are magnified by 10 for clarity.}
\label{harmonic}
\end{figure*}

Figure \ref{harmonic} shows the field $|\mathbf{g}_{M}|$ computed from equation (\ref{potential}), the field $|\mathbf{g}_{s}|$ computed from equation (\ref{naive}), and the MOND field computed numerically.  The numerical solution and the perturbative solution match well at small $\epsilon$.  The dashed line indicates the ratio of $|\mathbf{g}_{s}|$ to the numericaly computed $|\mathbf{g}_{M}|$.  It shows that the ratio diverges as the perturbation parameter $\epsilon$ approaches to zero.  This is indicative of the fact that $\mathbf{g}_{s}\simeq\frac{\mathbf{g}_{N}}{|\mathbf{g}_{N}|^{1/2}}\sim\epsilon^{\frac{1}{2}}$ whereas $\mathbf{g}_{M}\sim\epsilon$ inside the shell.  As a function of $\epsilon$, in MOND theory the perturbed field vanishes much more quickly than the naive expectation. 
\begin{figure*}[t!]
\centering{
(a)\includegraphics[width=3.0in]{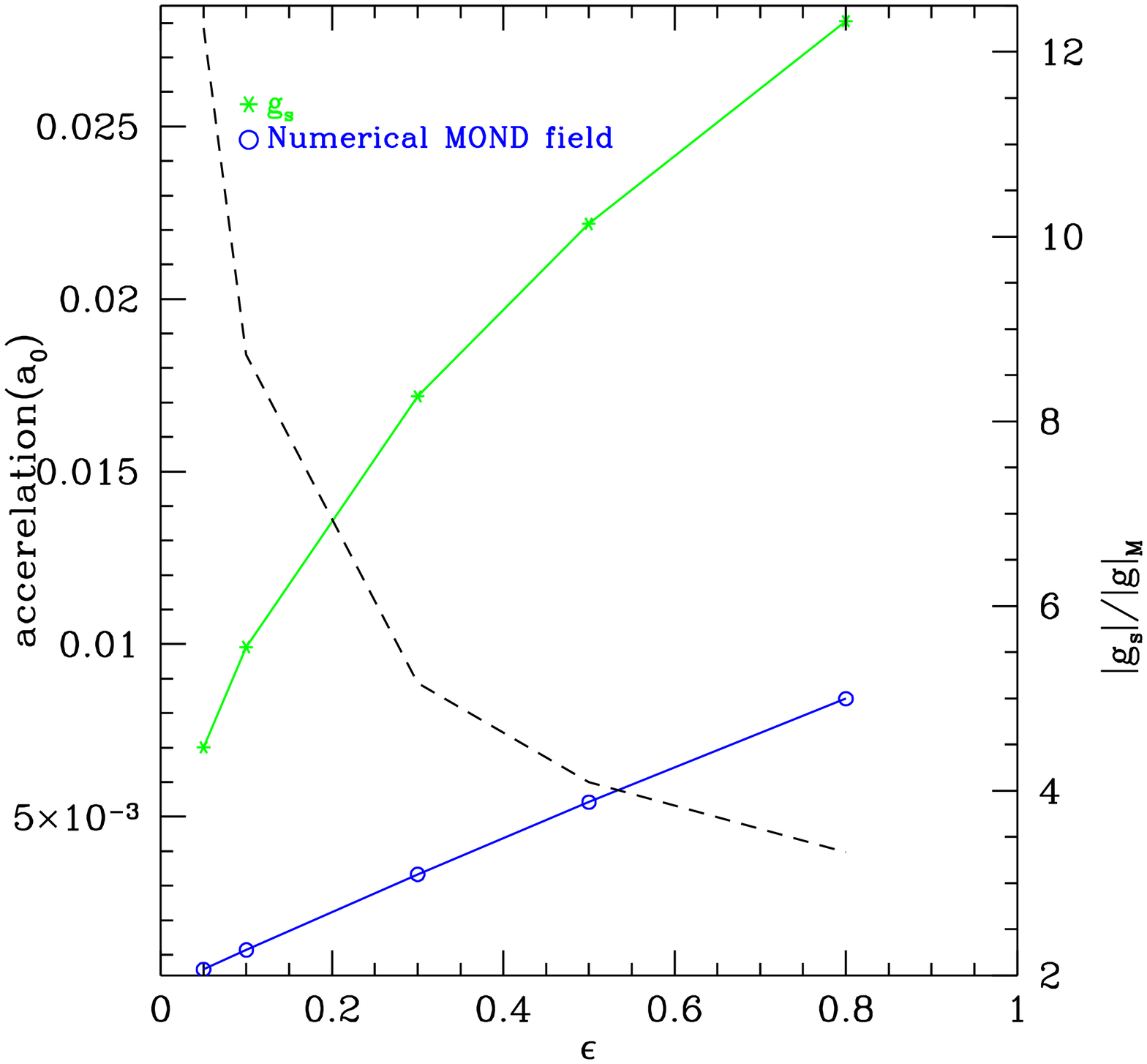}
(b)\includegraphics[width=3.0in]{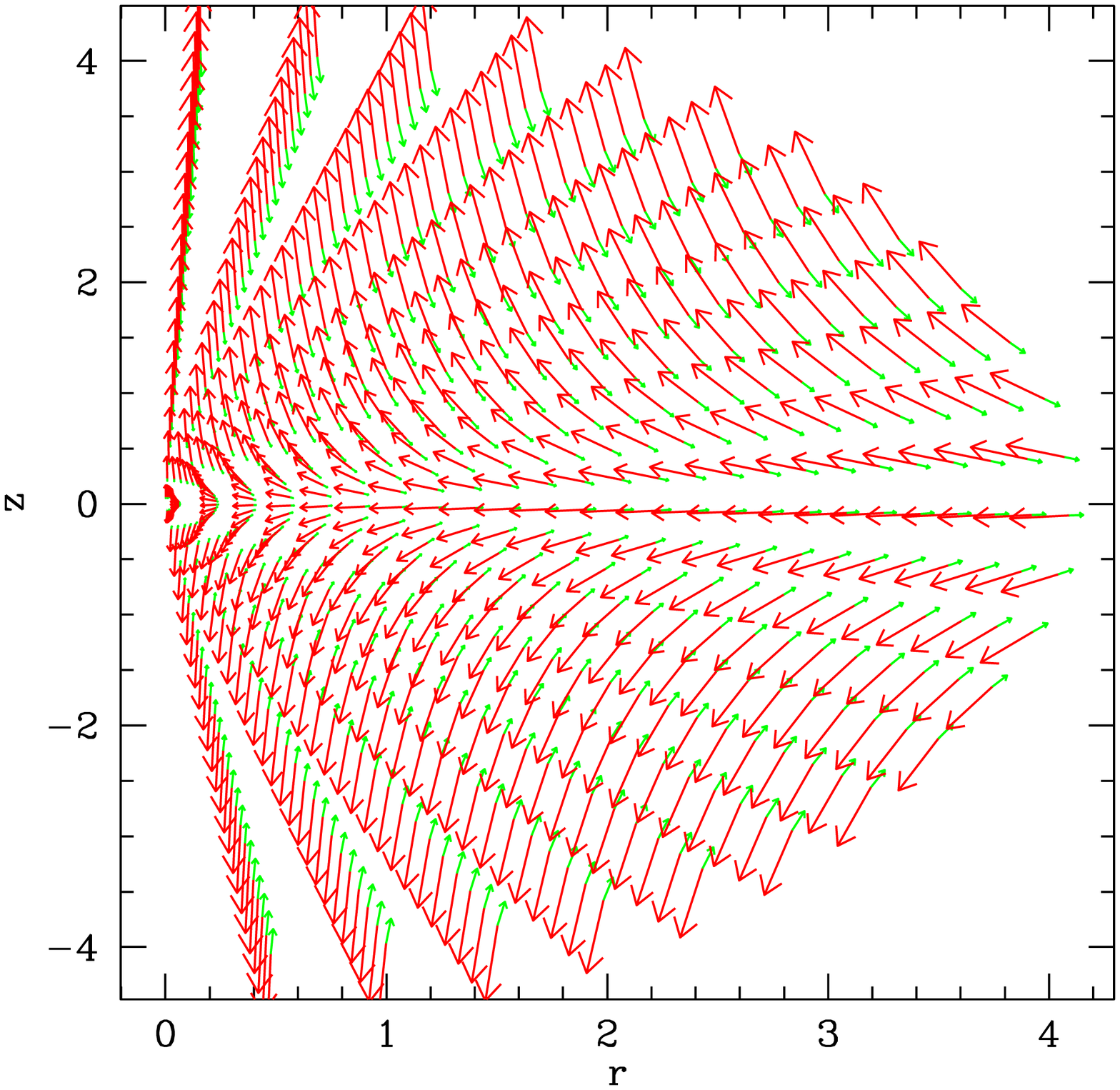}}
\caption{(a) This figure plots the field at a particular point  ($r=0.05R$ and $\theta=45^{\circ}$) inside a  shell
with an $l=2$ perturbation of the form given by equation (\ref{perturbation},
as a function of the strength $\epsilon$ of the perturbation.  The strength of the numerical MOND field $\mathbf{g}_{M}$, 
the perturbative MOND field and the naive MOND field $|\mathbf{g}_{s}|$ are shown. 
The dashed line indicates the ratio of $|\mathbf{g}_{s}|$ to $|\mathbf{g}_{M}|$ with the scale given on the right axis.  (b)This figure shows field configuration within the sphere for $l=2$ at $\epsilon=0.1$. Green arrows indicate the numerical MOND field, and red arrows indicate the difference of numerically calculated MOND field and the naive MOND field, $\mathbf{g}_{M}-\mathbf{g}_{s}$.  The size of the arrows are magnified by 10 for clarity}
\label{Twosphere2}
\end{figure*}  
\begin{figure*}[t!]
\centering{
\includegraphics[width=3.0in]{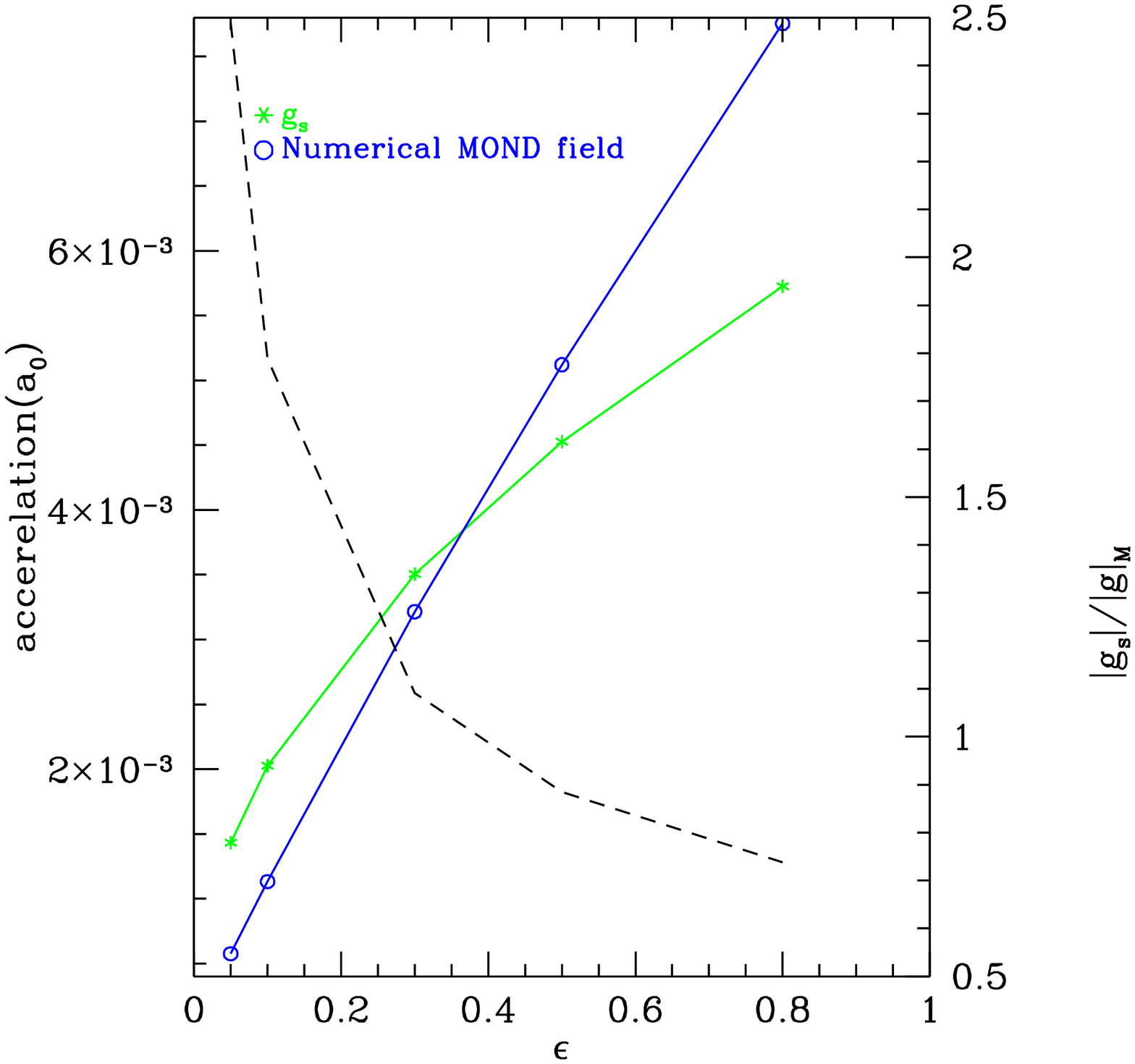}
\includegraphics[width=3.0in]{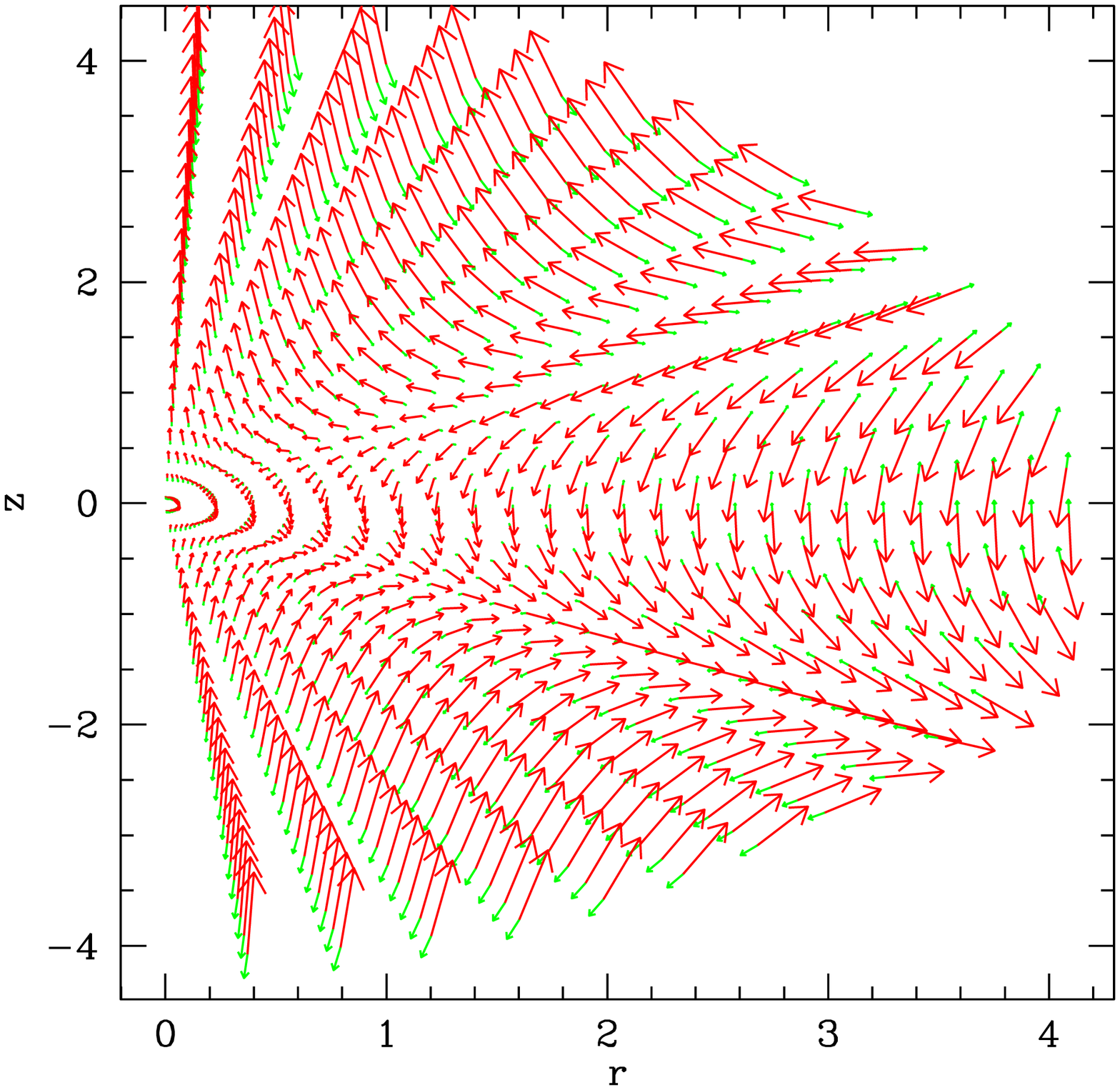}}
\caption{((a)This figure plots the field at a particular point  ($r=0.05R$ and $\theta=45^{\circ}$) inside a shell
with an $l=3$ perturbation  of the form given by equation (\ref{perturbation},
as a function of the strength $\epsilon$ of the perturbation.  The strength of the numerical MOND field $\mathbf{g}_{M}$, 
the perturbative MOND field and the naive MOND field $|\mathbf{g}_{s}|$ are shown. 
The dashed line indicates the ratio of $|\mathbf{g}_{s}|$ to $|\mathbf{g}_{M}|$ with the scale given on the right axis.  (b)This figure shows field configuration within the sphere for $l=3$ at $\epsilon=0.1$. Green arrows indicate the numerical MOND field, and red arrows indicate the difference of numerically calculated MOND field and the naive MOND field, $\mathbf{g}_{M}-\mathbf{g}_{s}$.  The size of the arrows are magnified by 10 for clarity}
\label{Twosphere3}
\end{figure*}  

For the higher order perturbation of $l\geq2$, the analytic solution of even the 'simple' Possion equation (\ref{simplemu}) is not known, and we rely entirely on the numerical simulation.  FIG. \ref{Twosphere2}(b) and FIG. \ref{Twosphere3}(b) show the field configuration inside the shell for $l=2$ and $l=3$ respectively.  Green arrows indicate the MOND field $\mathbf{g}_{M}$, and red arrows indicate the difference $\mathbf{g}_{M}-\mathbf{g}_{s}$.  From these figures, it is apparent that $\mathbf{g}_{M}<<\mathbf{g}_{s}$ inside the shell.  Similar to the $l=1$ case, the simple rescaling formula overestimates the field inside.  FIG. \ref{Twosphere2}(a) and FIG. \ref{Twosphere3}(a) shows $|\mathbf{g}_{M}|$ and $|\mathbf{g}_{s}|$ at a single point (r=0.05R and $\theta\simeq45^{\circ}$) as a function of perturbation parameter $\epsilon$.  As shown in the figures, just like $l=1$ case, the ratio $|\mathbf{g}_{s}|/|\mathbf{g}_{M}|$ diverges as $\epsilon\rightarrow 0$.  It indicates that $\mathbf{g}_{M}$ vanishes faster than $\mathbf{g}_{s}$ in the limit of small $\epsilon$.  

In general, the MOND field due to the small perturbation is suppressed compared to $\mathbf{g}_{s}$, 
and the aspherical perturbation external to the shell may not influence the internal system as much as one might naively expect.

\section{Shielding and Anti-Shielding effects}
Since the MOND equations are non-linear, it might be expected that changing the mass of the shell while keeping the perturbation term constant would affect the field inside the shell.  (In Newtonian dynamics, it does not.)  This is indeed the case.  Again, we consider a perturbation of the form equation (\ref{perturbation}) with $l=1$.  Keeping the $\sigma_{sh}=M_{s}/R^{2}$ term in equation (\ref{perturbation}) constant while increasing the mass of the shell from $M_{s}$ to $M_{new}$, the potential inside the shell becomes
\begin{eqnarray}
\psi_{1 new}&=&\sqrt{GM_{new}a_{0}}\frac{M_{sh}}{M_{new}}\epsilon\cos(\theta)\left[\frac{r}{2R}+\frac{6}{56}\left(\frac{r_{t new}}{R}\right)^{2}\frac{r}{R}\right] \nonumber \\
            &=&\Psi_{sh}\sqrt{\frac{M_{sh}}{M_{new}}}\epsilon\cos(\theta)\left[\frac{r}{2R}+\frac{6}{56}\left(\frac{r_{t new}}{R}\right)^{2}\frac{r}{R}\right] \nonumber \\
 \label{newpotential}
\end{eqnarray}

\begin{figure}
\centering{
\includegraphics[width=3.0in]{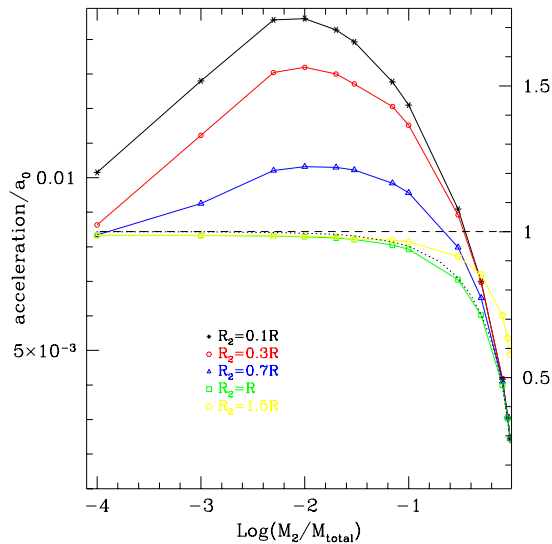}}
\caption{This figure shows the magnitude of field at  $r=0.05R$ and $\theta\simeq45^{\circ}$ as a function of $M_{2}/M_{total}$.  The mass and the radius of the first shell is fixed at $R=0.64 Mpc$ and $M_{sh}=2.0\times10^{13}M_{\odot}$.  The perturbation given in equation (\ref{perturbation}) with $l=1$ and $\epsilon=0.07$ is put on the first shell.  Each curve corresponds to the second shell having a different radius.  The horizontal dashed line indicates the MOND field of the single shell system with equivalent mass, radius and perturbation as the first shell.  The scale on the right is normalized to this value.  The dotted line represents the field corresponds to the perturbative solution of equation (\ref{newpotential}).}
\label{two}
\end{figure}

Comparing equation (\ref{newpotential}) and equation (\ref{potential}), equation (\ref{newpotential}) is suppressed by the factor of $\sqrt{\frac{M_{sh}}{M_{new}}}$.  Hence, the field within the shell is reduced by adding more mass to the shell.  The perturbed field within the shell is reduced by the spherically symmetric part of mass distribution.  This can be thought as a screening effect of the spherical shell.  As the spherically symmetric part 
of the mass distribution becomes more dominant, the perturbed field within the shell is reduced.  The field corresponds to equation (\ref{newpotential}), and is represented by the curve for $R_{2}=R$ in FIG. \ref{two}.  $M_{new}$ can be decomposed into $M_{sh}+M_{2}$ where $M_{2}$ is the spherically symmetric component that has been added.

The situation described in equation (\ref{newpotential}) can be thought as a special case of a two-shell system where the radius of the first shell R and that of the second shell $R_{2}$ coincide.    
FIG. \ref{two} depicts the general case where the two radii are different.                    
Five curves are plotted corresponding to different radii for the second shell.  The horizontal dashed line indicate the MOND field of the first shell with the perturbation when the second shell is not present.   In the limit where  $M_{2}$ vanishes, the field will reduce to this value.  From FIG. \ref{two}, one can observe that, for a generic value of the inner mass ($M_{2}\sim M_{sh}$), the field is screened regardless of the radius of the second shell.  Increasing the mass of the spherically symmetric part of distribution, normally screens the field from the aspherical part of the mass distribution.  The screening will be most efficient for the special case when the two shells coincide (i.e. $R=R_{2}$), and in this case field inside is always suppressed. 
 When $R_{2}>R$, the field is again always suppressed.  For a given value of $M_{2}$, the suppression factor is less than that of the special case $R=R_{2}$.  The case $R_{2}<R$ exhibits a carious feature.  The field inside the second shell can be enhanced when $M_{2}<<M_{sh}$, this is an anti-screening effect.  We also note that this enhancement is larger for $R_{2}<<R$.  

When $M_{2}<<M_{sh}$ and $R_{2}<<R$, the second shell belongs to the local system while the first shell is the background mass distribution. The local shell then picks up and enhance the perturbations of the background mass distribution.
 \begin{figure}
\centering{
\includegraphics[width=3.0in]{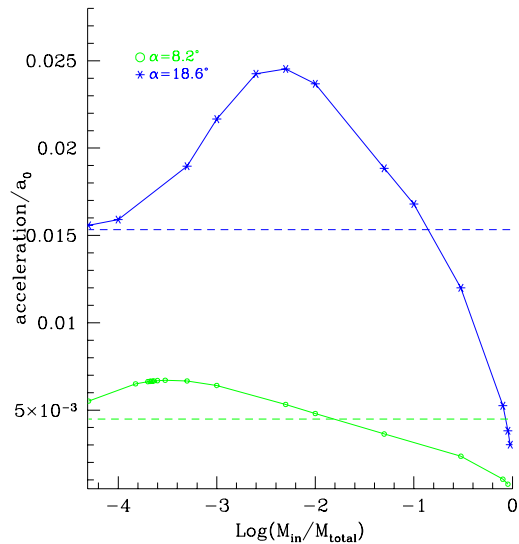}}
\caption{This figure shows the magnitude of the field at  $r=0.05R$ and $\theta\simeq45^{\circ}$ as a function of $M_{2}$.  The radius of the second shell is fixed at $R_{2}=0.3R$.  The perturbation on the outer shell is provided by an opening on top of the shell.  Two curves are shown for the opening angle of $18.6^{\circ}$ (blue) and $8.2^{\circ}$ (green).  The dashed line indicates the corresponding field without the presence of the inner shell.}
\label{bowl}
\end{figure}  

Finally, we consider the two-shell system with a perturbation other than pure $l=1$.  Specifically, a spherical cap is removed from the first shell with opening angle $\alpha$, measured from positive z-axis, leaving behind a spherical bowl.  The mass density of the bowl is increased so as to keep  the total mass of the bowl equal to that of the original full shell.   We consider opening angles of $8.2^{\circ}$ and $18.6^{\circ}$.  A perturbation of this form contains the sum of Legendre Polynomials up to infinite oreder.  

Although the form of the perturbation is very different, FIG.\ref{bowl} shows much the same feature as FIG.\ref{two}. For $M_{2}\sim M_{sh}$ the field is screened.  For larger breaking of the spherical symmetry (larger $\alpha$), the screening becomes less effecient and it requires more mass for the second shell to suppress the field.  For $M_{2}<<M_{sh}$, the field within the second shell is again enhanced.  It might be expected that the enhancement becomes prominent when the generic MOND field of the second shell $\sim\sqrt{GM_{2}a_{0}}/R_{2}$ becomes comparable to the field produced by the bowl-only system.  For example, for the opening angle of $\alpha=8.2^{\circ}$, the bowl produces an average MOND field of $\sim 0.0052a_{0}$ at $r=R_{2}$.  The generic MOND field of the second shell becomes comparable when $Log[M_{2}/M_{total}]\sim -2.1$.  Similarly, for the opening angle of $\alpha=18.6^{\circ}$, the MOND field of the second shell becomes comparable at $Log[M_{2}/M_{total}]\sim -1.6$.  These values reasonably estimate the mass when the enhancement starts to become important.       

\section{Concluding remarks}            
We have shown, by examining a simple toy system of a shell with a perturbation that the naive MOND equation (\ref{naive}) overestimates the field inside the shell.  For a small perturbation on the shell signified by the perturbation paramter $\epsilon$, the field within the shell shell is of order $\epsilon$ and not $\epsilon^{\frac{1}{2}}$ as predicted by the naive MOND equation.  To properly estimate the field within the shell, the jump condition on the surface of the shell has to be taken into account.
We have also shown that the field due to the perturbation within the shell can be screened by the spherical distribution of the mass.  The screening is strongest  if the spherical distribution is positioned near the shell.  This contrasts to Newtonian gravity where addition of the spherical distribution does not affect the field configuration within the distribution.
Finally, the anti-screening effect is noticed when the small and light shell is added to the system.  The enhancement inside the small shell becomes significant when the generic field of the shell $\sim\sqrt{GMa_{0}}/R$ becomes comparable to the perturbation field near the shell.  The enhancement is generally larger for the shell with smaller radius.

Let now suppose that, in the simplest picture, a cluster is modeled by a spherical shell and the external field is produced by an aspherical perturbation on the shell.  In order to estimate the magnitude of the external field, one needs to know not only the size of perturbation but also the total mass of the shell.  Applying naive MOND equation on this system to estimate the external field would be quite inadequate.  If the galaxy can be modeled by a small shell within a large, more dominant shell (a cluster), then the external field within the galaxy is likely to be enhanced.  The enhancement depends on the ratio $M_{gal}/M_{clu}$ and the size of the galaxy, but as shown in FIG. \ref{two}, it can be of the same size as the external field itself.                
\section{acknowledgements}
GDS and RM were supported by a grant from the US Department of Energy to the particle astrophysics theory group at  CWRU.

\end{document}